# A source of entangled photons based on a cavity-enhanced and strain-tuned GaAs quantum dot

Michele B. Rota[1]*, Tobias M. Krieger[2]*, Quirin Buchinger[3], Mattia Beccaceci[1], Julia Neuwirth[1], Hêlio Huet[1], Nikola Horová[4], Gabriele Lovicu[1], Giuseppe Ronco[1] Saimon F. Covre da Silva[2], Giorgio Pettinari[5], Magdalena Moczała-Dusanowska[3], Christoph Kohlberger[2], Santanu Manna[2], Sandra Stroj[6], Julia Freund[2], Xueyong Yuan[2,7], Christian Schneider[8], Miroslav Ježek[4], Sven Höfling[3], Francesco Basso Basset[1], Tobias Huber-Loyola[3+], Armando Rastelli[2+]
& Rinaldo Trotta[1+]

[1]Dipartimento di Fisica, Sapienza University of Rome, Piazzale Aldo Moro 5, I-00185 Rome, Italy.

[2]Institute of Semiconductor and Solid State Physics, Johannes Kepler University,

Altenberger Strasse 69, A-4040 Linz, Austria.

[3]Technische Physik, Universität Würzburg, Am Hubland, D-97074 Würzburg, Germany.

[4]Department of Optics, Faculty of Science, Palacký University, 17. listopadu 1192/12, 77900 Olomouc, Czech Republic.

[5]Institute for Photonics and Nanotechnologies, National Research Council, Via del Fosso del Cavaliere, 100, 00133 Roma, Italy.

[6]Forschungszentrum Mikrotechnik, FH Vorarlberg, Hochschulstr. 1, A-6850 Dornbirn, Austria.

[7]School of Physics, Southeast University, 211189 Nanjing, China.

[8]Institute of Physics, University of Oldenburg, D-26129 Oldenburg, Germany

*These authors contributed equally to this work.

+Corresponding authors' emails:

tobias.huber@physik.uni-wuerzburg.de

armando.rastelli@jku.at

rinaldo.trotta@uniroma1.it

## Abstract
A quantum-light source that delivers photons with a high brightness and a high degree of entanglement is fundamental for the development of efficient entanglement-based quantum-key distribution systems. Among all possible candidates, epitaxial quantum dots are currently emerging as one of the brightest sources of highly entangled photons. However, the optimization of both brightness and entanglement currently requires different technologies that are difficult to combine in a scalable manner. In this work, we overcome this challenge by developing a novel device consisting of a quantum dot embedded in a circular Bragg resonator, in turn, integrated onto a micromachined piezoelectric actuator. The resonator engineers the light-matter interaction to empower extraction efficiencies up to 0.69(4). Simultaneously, the actuator manipulates strain fields that tune the quantum dot for the generation of entangled photons with corrected fidelities to a maximally entangled state up to 0.96(1). This hybrid technology has the potential to overcome the limitations of the key rates that plague QD-based entangled sources for entanglement-based quantum key distribution and entanglement-based quantum networks.

## Introduction
Scalable sources of entangled photons are the keystone for the realization of a photonic quantum network[1–5] where quantum bits of information are, for example, encoded in the polarization state of single photons and travel between different nodes of the network[6]. To date, the majority of entanglement-based quantum communication protocols have been implemented using sources based on spontaneous parametric down-conversion (SPDC) processes[7,8]. SPDC sources can generate high-fidelity entangled photons with high brightness and provide the possibility of exploiting entanglement on different degrees of freedom[9,10]. However, they are fundamentally limited by their probabilistic emission[11,12] which can reduce the maximal rate of operation and hinder scaling up to large photon number applications[9].

Quantum emitters driven under resonant excitation have instead the potential to overcome these hurdles: because of the Pauli exclusion principle and Coulomb interaction, each excited state can be populated only once, and the simultaneous emission of more than one photon of a given frequency per excitation cycle is reduced to the negligible probability of re-excitation during the same laser pulse[13,14]. Moreover, the use of a resonant excitation scheme with near-unity preparation fidelity[15–17] opens the possibility of achieving on-demand photon generation.

Among the plethora of quantum emitters available to date, e.g., colour centres in diamond[18], defects in 2D materials[19,20], semiconductor quantum dots (QDs) are arguably the most promising sources of entangled photons[21,22]. They can generate pairs of photons[23] on demand[15–17], with high photon flux[24–28], high indistinguishability[16,27,29], and high entanglement fidelity[30–33], and their emission properties can be tailored by adjusting the growth parameters[34] and/or by the application of external perturbations[35–38]. In the last few years, proof-of-concept experiments, such as quantum teleportation[39,40], entanglement swapping[41,42], generation of cluster states[43,44], and entanglement-based quantum key distribution[45,46], have demonstrated the potential of QD-based entangled photon sources. Despite these accomplishments, their exploitation in real-life applications is still in its infancy. The main reason is that applications demand the simultaneous optimization of several different figures of merit of the source. However, to date, each of them requires advanced technological solutions that are often incompatible with each other. To provide an example, let us consider a point-to-point entanglement-based quantum key distribution[47]. For this application, particularly for the implementation of device-independent scenarios[48], it is fundamental to minimize the quantum bit error rate and simultaneously maximize the key rate. Looking at the photon source, this prospect implies simultaneously boosting to near-unity values (*i*) the degree of entanglement and (*ii*) the photon extraction efficiency – a task that is far from easy for QDs, as detailed below.

Concerning (*i*), QDs can generate pairs of polarization-entangled photons via the radiative cascade of a biexciton (XX) to the ground state (0) via the intermediate exciton (X) levels[49]. Experiments have demonstrated that the measured degree of entanglement can be limited by several physical processes, including hyperfine interaction[50], optical Stark effect[51], re-excitation[24], exciton scattering with excess charges[52], and anisotropic electron-hole exchange interaction[53]. After 20 years of research on the subject, a degree of entanglement as high as 0.98 was finally achieved[30], without resorting to inefficient and impractical temporal/spectral post-selection. The key ingredients are the use of GaAs/AlGaAs QD samples with short radiative decay times[54], two-photon resonant excitation[15,17], and, most importantly, anisotropic strain fields delivered by multi-axial piezoelectric actuators[55,56]. The latter can be used to cancel any residual fine structure splitting (FSS, with a magnitude *s*) between the intermediate X states (induced by the anisotropic electron-hole exchange interaction), which leads to the evolution of the entangled state over time, which cannot be fully captured by detectors with finite time resolution. Several external perturbations (such as strain, electric, or magnetic fields[57] or a combination of them) can be used to erase the FSS; our choice is to rely on multi-axial strain fields only which, until now, is the only method that has demonstrated the capability to achieve a near-unity entanglement degree by erasing the FSS virtually in any QD in the sample.

Concerning (*ii*), embedding a single QD inside an optical cavity[58] is a common strategy adopted to increase the photon extraction efficiency, which is reduced by total internal reflection at the semiconductor-vacuum interface[59], as it allows coupling of the QD emission into the tailored far-field emission pattern of a single mode of the electromagnetic field. This approach also enables the acceleration of spontaneous emission via the Purcell effect[60], opening the path towards GHz operation rates[27]. Over the years, a variety of optical cavities have been used to enhance single optical transitions in QDs[58], with open cavities-systems[61] currently setting the state of the art for single-photon sources. More sophisticated approaches must instead be used for entangled photon sources, mainly because the energy of the photons generated in the XX cascade features a difference in energy[62] (a few meV, due to Coulomb interaction among the carriers). Thus, researchers have resorted to photonic molecules[24], nanowires[25], dielectric antennas[26], low-Q micropillars[63], as well as circular Bragg resonators (CBR) or bullseye cavities[64]. In particular, recent works on QDs embedded in CBRs demonstrated extraction efficiencies as high as 0.85(3)[27] with the highest reported entanglement fidelity of 0.90(1) without any reduction of the residual FSS[28].

Merging the CBRs and multiaxial-strain-tuning technologies would be the ideal choice for applications in the field of quantum key distribution. However, this task turned out to be technologically challenging because of the need to attain tight control over the anisotropy of the strain transferred to the QDs embedded in the CBR cavities. Previous attempts were limited to the use of monolithic piezoelectric substrates which cannot be used to tailor the in-plane strain anisotropy and are therefore not suitable for the erasure of FSS and the generation of highly entangled photons[65]. In this work, we overcome these obstacles and present the first entangled-photon emitter based on a single GaAs/AlGaAs QD, embedded in a CBR, and integrated onto a micro-machined piezoelectric actuator that allows for three-axial strain engineering. This device combines at the same time high brightness, energy tuning, and entanglement optimization.

# Results

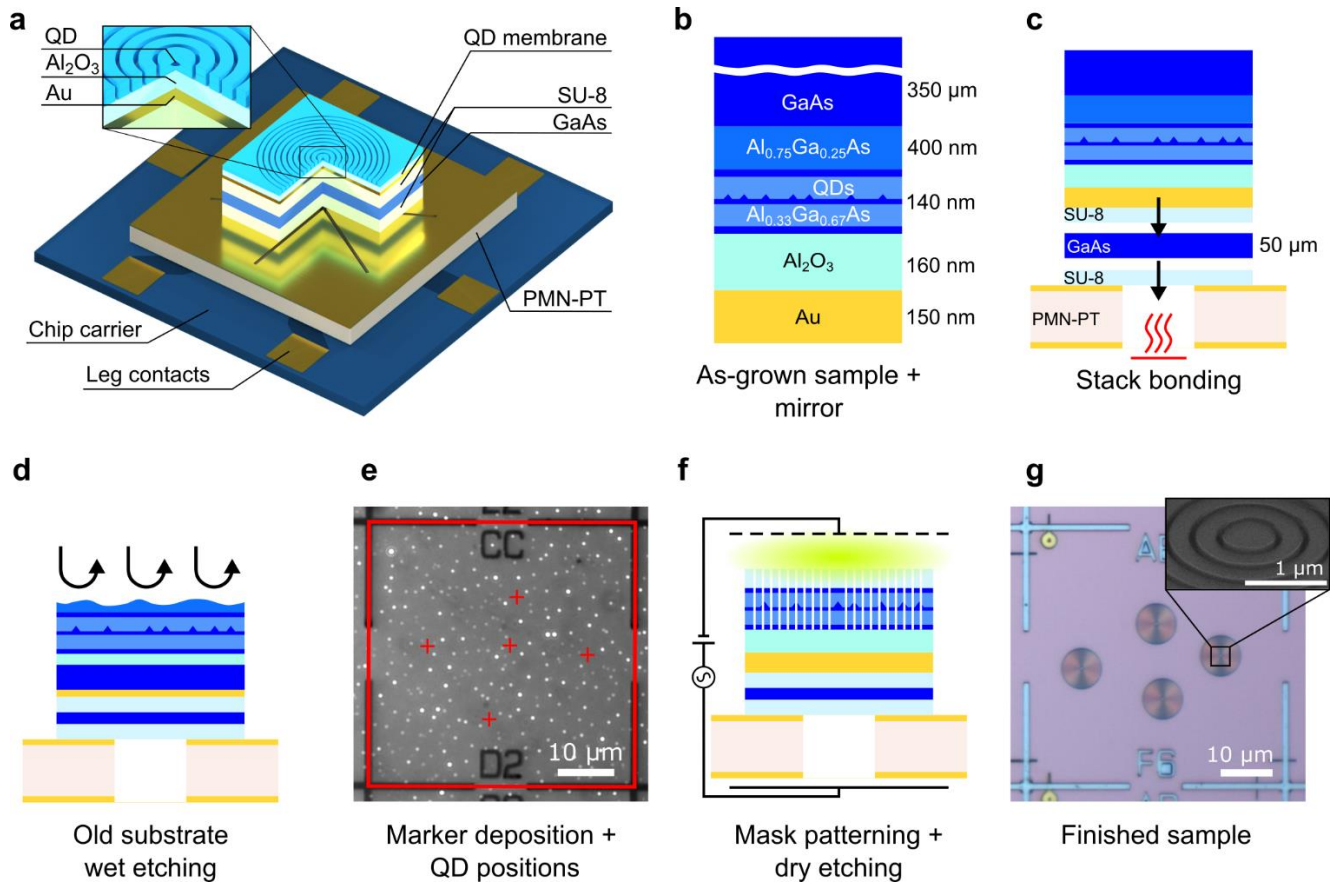

**Figure 1 | Processing steps for fabricating a circular Bragg resonator (CBR) cavity on a piezoelectric substrate.** **(a)** Schematic of a CBR sample on six-legged piezoelectric substrate mounted on a chip carrier. Dimensions are not to scale. **(b)** As-grown quantum dot (QD) sample structure with the oxide and metal mirror deposited on the surface. **(c)** The sample is bonded with SU-8 photoresist on a GaAs carrier by applying pressure and heat to reach the curing temperature of the photoresist (230 °C). The carrier is later lapped to reduce its thickness to approximately 50 µm. After thinning, the sample is bonded suspended onto the six-legged piezoelectric substrate by using the same procedure. **(d)** The original substrate and the sacrificial layer are removed via wet etching. **(e)** Cryogenic optical microscope image showing the photoluminescence of single QDs and a square grid of metallic markers defined on the sample surface via electron beam lithography (EBL) and metal deposition to create a frame of reference. The red square is the result of marker recognition obtained with image processing software. The red crosses represent the positions of single QDs obtained with a 2D Gaussian fit of the QD emission. **(f)** CBRs are defined in a second EBL step around preselected single QDs. The masked sample is then dry-etched with chlorine and argon plasma in an inductively coupled plasma machine to transfer the cavities onto the membrane. **(g)** Optical microscopy image of a finished sample. A tilted scanning electron microscope image of the centre of a single structure is shown in the inset.

**Sample Processing.** A sketch of the device discussed in this work is shown in Fig. 1a. The cavity consists of a ~670 nm disc with the QD in its geometrical centre, surrounded by a circular Bragg grating made of several trenches with a period that matches the second-order Bragg condition[66] to reflect light travelling inside the semiconductor in the orthogonal direction. The grating, combined with a metallic mirror below the structure, results in a quasi-Gaussian emission profile from the top surface. The CBR features a rather flat extraction efficiency profile over tens of nm of wavelength and a low-Q (~100) cavity mode enabling modest Purcell enhancement of both the X and XX emission[67–69]. The entire cavity is integrated onto a 300 μm thick $[Pb(Mg_{1/3}Nb_{2/3})O_3]_{0.72}$-$[PbTiO_3]_{0.28}$ (PMN-PT) piezoelectric plate micromachined via femtosecond-laser cut into six different actuators ("legs") aligned at 60° to each other. Voltages applied to pairs of aligned legs control three independent strain fields that can be used to tune QDs for the generation of entangled photons with tuneable energy[55]. The integration of the CBR cavity onto the micromachined piezoelectric actuator requires several different technological advances, as discussed in more detail below.

The processing starts with the sample grown on a GaAs(001) substrate in a molecular beam epitaxy (MBE) machine (see Methods for further details). The optically active area consists of a layer of GaAs QDs obtained via the Al-droplet-etching-epitaxy technique[54] placed at the centre of a 140 nm thick $Al_{0.33}Ga_{0.67}As$ layer sandwiched between two 4-nm thick GaAs protective layers. An $Al_{0.75}Ga_{0.25}As$ layer is grown below the active layer to enable substrate removal.

The surface of the sample is coated with a broadband mirror in a two-step deposition process, with a gold film deposited on top of an aluminium oxide layer to avoid potential plasmonic losses at the metal interface[70], as shown in Fig. 1b.

The Au-coated surface is then bonded with a photoresist (SU-8) to a carrier made of a 350 μm thick GaAs substrate by the application of pressure and high temperature (230 °C). The GaAs carrier is thinned by mechanical lapping to a final thickness of less than 50 μm. The thickness of the GaAs carrier must be as low as possible to ensure the largest strain transfer[55] while providing reliable mechanical support to the membrane during all processing steps. The sample with the thinned-down carrier is bonded with SU-8 to the micromachined piezoelectric substrate, as shown in Fig. 1c. The contacts on the piezoelectric substrates allow the application of three independent voltages on pairs of aligned legs, as described elsewhere[36]. The original GaAs substrate, together with the sacrificial layer, is removed with a three-step wet etching procedure, as shown in Fig. 1d. The surface of the resulting QD nanomembrane on top of the oxide/Au reflector is spin-coated with an electron-beam resist and patterned using electron beam lithography (EBL) with a square grid of markers. A 150 nm thick stack of equally thick (strain-compensated) Cr-Au-Cr layers is evaporated onto the surface, forming the grid after lift-off. The markers are used as a frame of reference to acquire the positions of single QDs with 15 nm precision in a cryogenic microscopy setup[71], see Fig. 1e. The acquired positions are used to create a pattern design of the microcavities with single QDs at their centres in a second EBL step. The patterned cavity designs are transferred onto the semiconductor membrane by dry etching in an inductively coupled plasma machine (Fig. 1f), followed by mask removal (Fig. 1g). Given the narrow emission energy distribution of the QDs, the broadband response of the cavities, and tunability of the final device, the cavity design is adjusted to match the average emission energy of the QD ensemble without adapting the design of each cavity to the properties of the embedded emitter. For complete details about all the processing procedures and parameters and the QD position mapping method, see section S1 of the Supplementary Information.

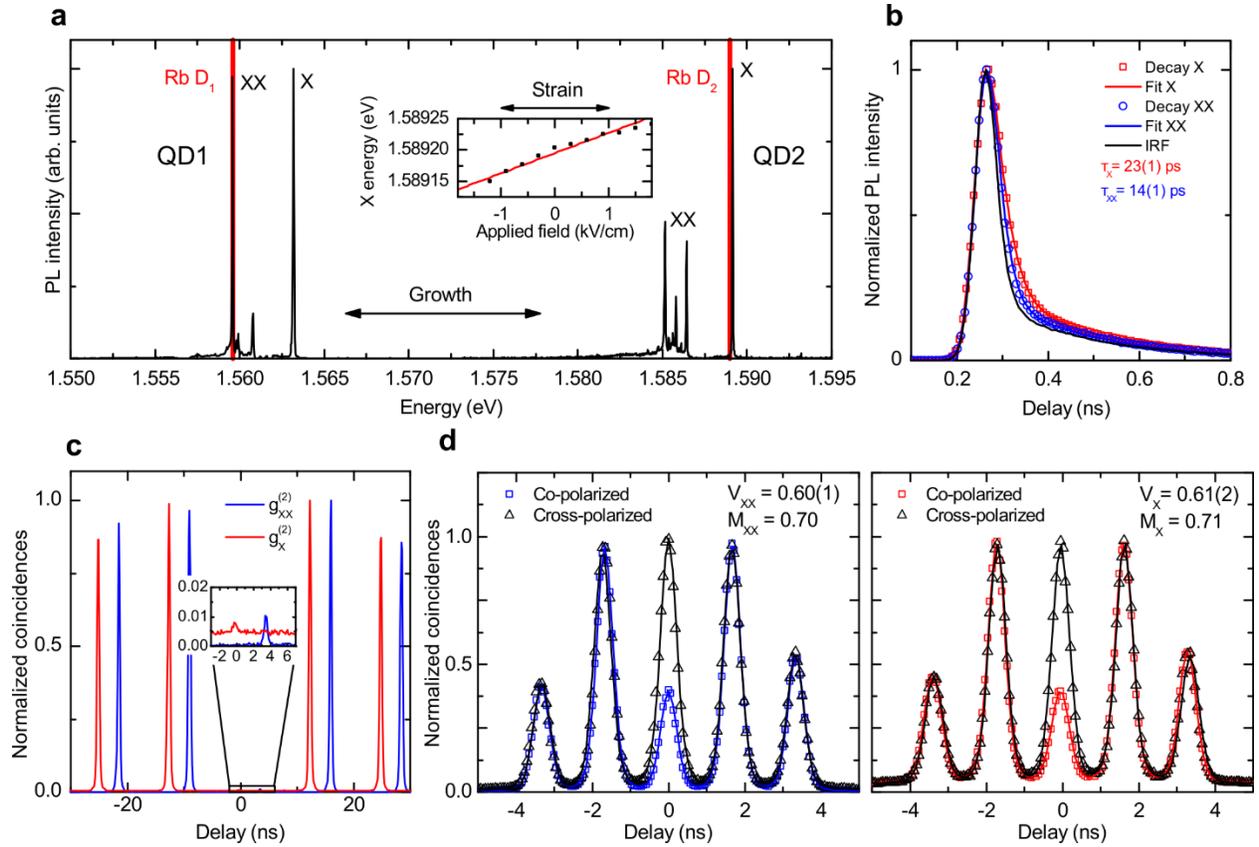

**Figure 2 | Optical characterization of cavity-enhanced QDs. (a)** Photoluminescence spectra of two representative QDs (labelled QD1 and QD2) from two different samples showing emission in the vicinity of the $D_1$ and $D_2$ transitions of rubidium (Rb). The exciton (X) and biexciton (XX) transitions are labelled. The emission of the QD can be tuned toward the resonance of the Rb transitions with the application of stress, as shown in the inset. **(b)** Time-resolved traces of the X (red squares) and XX (blue circles) transition intensities from another QD (QD3) and instrument response function (IRF) (black solid line) of the setup. The lifetime values are obtained with a fit (solid lines) convoluting the IRF with the exponential decay functions expected from the radiative cascade. **(c)** $g^{(2)}(\tau)$ histograms of the X (red line) and XX (blue line) emission lines for QD2. The histograms are shifted horizontally for ease of reading. The graphs around the 0-time delay are magnified in the inset to highlight the residual low coincidences. The values of the $g^{(2)}(0)$ are $g^{(2)}_{XX}(0) = 0.012(1)$ and $g^{(2)}_{X}(0) = 0.016(1)$. **(d)** Histograms of Hong-Ou-Mandel interference between co- (red squares) and cross-polarized (black triangles) photons from XX and X decay from QD2. The values of the visibility V are obtained using Gaussian peaks convoluted with an exponential decay fit of the peaks (solid lines). The values for the indistinguishability $M_X = 0.71$ and $M_{XX} = 0.70$ are calculated by considering the imperfections of the setup and the values of the $g^{(2)}(0)$, see text for more details.

**Optical Characterization**. Fig. 2a shows the low-temperature (5 K) spectra of two exemplary

QDs from two different samples excited under resonant two-photon excitation (TPE)[15,17] with a laser pulse length of 5(1) ps (see Methods). The two QDs emit in the vicinity of the $D_1$ and $D_2$ transitions of rubidium (Rb), a possible material choice for the realization of quantum memories[36,72]. The two most intense lines in the two spectra correspond to the XX and X transitions, as indicated for both the low-energy QD (QD1) and the high-energy QD (QD2). The other lines and the broad band below the XX transitions are likely due to other charged excitonic states[73]. Fine-tuning of the emission energy toward resonance with the Rb transitions can be achieved by applying voltages to one of the legs of the actuator (see the inset of Fig. 1a). The average tuning range achieved for the devices discussed in this work is about 90 neV/V.

To evaluate the acceleration of the spontaneous emission due to the Purcell effect we collect time-resolved emission decay traces. They are shown in Fig. 2b for the X and XX transitions of another QD (QD3), featuring an X emission energy of 1.589 eV, together with the instrument response function (IRF). A fit to the experimental data provides lifetimes of 23(1) ps and 14(1) ps for the X and XX transitions, respectively. Considering the measured lifetimes in the bulk sample[41], we calculate a Purcell factor of 11.7(5) and 9.3(5) for XX and X, respectively.

One of the most important properties of a quantum emitter is the ability to emit only a single photon in a given spectral window per excitation pulse. To evaluate the probability of multiphoton emission we perform autocorrelation measurements and estimate the value of the second-order correlation function $g^{(2)}(\tau)$ at $\tau = 0$. For these measurements, the duration of the excitation laser pulse was set to 1.9(3) ps to limit the effect of re-excitation during the same laser pulse[14]. The $g^{(2)}(\tau)$ histograms for QD2 are shown in Fig. 2c for both the X and XX photons, red and blue curves, respectively. The values obtained from the histograms, $g^{(2)}_{XX}(0) = 0.012(1)$ and $g^{(2)}_{X}(0) = 0.016(1)$, are mainly limited by the use of a white halogen lamp to mitigate blinking[74] and by the non-perfect rejection of the resonant laser and/or QD sidebands (see Methods).

The extraction efficiency, i.e., the fraction of photons collected by the lens on top of the samples for the X and XX photons are $\eta_X = 0.67(3)$ and $\eta_{XX} = 0.69(4)$. For a laser with a repetition rate of 80 MHz, these results in a measured 3.13(1) Mcps and 3.52(1) Mcps at the single photon avalanche photodiodes (SPADs) for the brightest QD in the sample, resulting in 0.155(1) Mcps of measured X-XX coincidences. Considering the efficiency and the nonlinear response (due to the dead time) of the detectors, we estimated an average rate of single photons that arrive at each detector of 9.6(1) Mcps. The pair brightness at the first lens, i.e., the amount of photon pairs that arrive at the first lens of the setup per excitation pulse, is 0.13(1). Is calculated by multiplying the pair emission efficiency $\eta_{pair} = 0.279(3)$, which contains all the effects reducing the pair emission rate, i.e., the preparation fidelity and the blinking of the QD, times both the extraction efficiencies $\eta_X$ and $\eta_{XX}$. For more details on the calculation of the different figures of merit see section S5 of the Supplementary Information.

Another important property of the emitted photons is their indistinguishability, i.e., the degree of similarity between subsequently emitted photons in terms of their different degree of freedom, such as energy, dispersion, and wavepacket shape. Indistinguishability is fundamental in all applications that need the interference of two photons, e.g., quantum teleportation and entanglement swapping, as the fidelity of the process decreases steeply as the indistinguishability of the two involved photons decreases[75]. To assess the indistinguishability of the emitted photons, we measure the two-photon interference visibility by exploiting the Hong-Ou-Mandel effect[76] (see Methods). A histogram of the coincidences between the two exit ports of the beam splitter, where photons emitted are allowed to interfere, is shown in Fig. 2d for QD2. The photon states are prepared before interference with both the same and orthogonal polarization. The visibility of the 0-time delay peak can be used to calculate the indistinguishability of the emitted photons[77] and it is obtained from the data with a fit model made by Gaussian functions convoluted with an exponential decay curve. The visibility values obtained from the fit are 0.60(1) and 0.61(2) for the

XX and X photons, respectively. It is worth mentioning that these values are obtained without resorting to any spectral or temporal filtering of the photons. By considering the values for the visibility of the interferometer (0.96), the non-zero values of the $g^{(2)}(0)$ (0.025(5) obtained under similar conditions for both X and XX photons), and the non-ideal beam splitter ratio (R=0.48), we calculate[77] an indistinguishability $M_X = 0.71$ and $M_{XX} = 0.70$. These values are mainly limited by the time-correlations between the two photons emitted during the cascade. The theoretical upper limit[29], which depends on the ratio between the lifetimes of the XX and X, is 0.71(2) for this particular QD, featuring a transition lifetime of 44(3) ps for X and 18(1) ps for XX (see section S3 of the Supporting Information for lifetime data), which is in excellent agreement with the measured values.

**Strain Tuning and Entanglement Recovery.** To recover the maximum entanglement degree of the two emitted photons when measuring with detectors with finite time resolution, the FSS of the QD must be reduced to a value much smaller than the natural linewidth of the emission[49]. In the case of Purcell-enhanced emission, the increase of the natural linewidth of the emitted photons (due to the accelerated spontaneous emission rate) strongly relaxes the demand for an ultra-small FSS. As an example, the natural linewidth of the X transition changes from 2.4 µeV for a 270 ps lifetime (typical for QDs in as-grown samples[41]) to 15 µeV for a 40 ps lifetime[49] (i.e., for a Purcell factor of about 7, easily achievable for QDs in our device). Therefore, we expect that QDs featuring large Purcell enhancement will generate entangled photons already at relatively large FSS. To observe this effect, we adjust the voltage applied to the micromachined piezoelectric actuator to restore the in-plane symmetry of the QD confining potential[78] and tune the FSS below the resolution of our set-up, while measuring the entanglement of the emitted photons. As described in previous works[55], this task can be accomplished by sweeping the voltages of one pair of legs for different values of the voltages applied to another pair of legs (the third pair of legs can be used to change the energy of the emitted photons at zero FSS). As shown in Fig. 3a for

QD2, this allows us to quickly identify the unique[36] set of electric fields applied to the piezoelectric that allows suppressing the FSS, being $E_{1-4}$=12 kV/cm and $E_{2-5}$=6.67 kV/cm, see also section S6 of the Supplementary Information. The procedure is better understood by looking at the polarization dependence of the emission energy of the X and XX. The polar plots in Figs. 3b and 3c report the polarization dependence of the deviation of the X emission from its unperturbed value. The amplitude of the sinusoid is the magnitude of the FSS while the phase gives the polarization direction of the X emission. By applying an electric field on legs 2 and 5 while keeping legs 1 and 4 at 0 kV/cm, we apply stress to the QD (straight arrows) and rotate the polarization axis (curved arrows) until it is aligned with the direction along which the stress is applied by another pair of legs, see the green and the dark orange plots in Fig. 3b, corresponding to the points in Fig. 3a circled with the same colour. After this, we change the electric field on the now aligned legs, e.g., 1 and 4, see the dark orange points in Fig. 3c, until the oscillation of the emission energy goes to zero, dark blue points in Fig 3c. In this condition, the emission energy of the QD does not depend anymore on the selected polarization since the degeneracy of the X level is restored, i.e., the FSS is erased, corresponding to the dark blue circled point in Fig 3a.

To measure the entanglement of the photon pair and gain complete information on its polarization state, we performed a quantum state tomography of the X-XX two-photon state for different decreasing values of the FSS down to zero. By reconstructing the two-photon density matrix, we can estimate the degree of entanglement in terms of the maximal fidelity to a Bell state that can be achieved with simple unitary transformations. This quantity is also known as fully entangled fraction[79] (FEF) and is not affected by possible undesired rotations in the polarization states. It is defined as the overlap between the experimental state $\hat{\rho}$ and a maximally entangled state $|\Phi\rangle$ maximized over all possible choices of $|\Phi\rangle$: $F = \max_{|\Phi\rangle}\{\langle\Phi|\hat{\rho}|\Phi\rangle\}$.

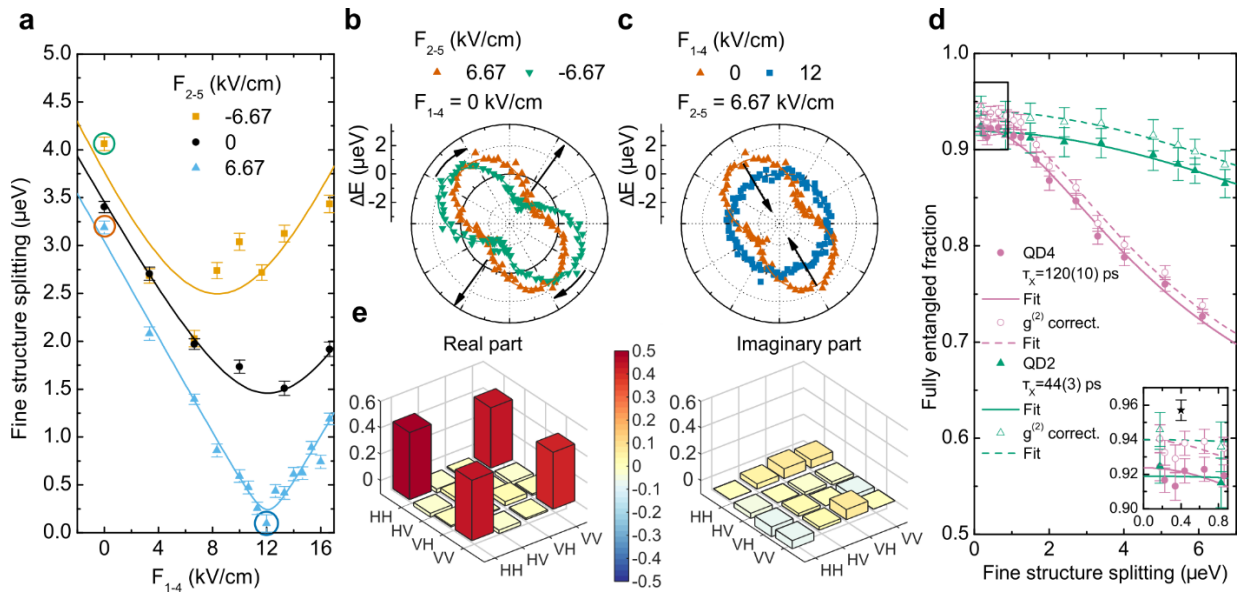

**Figure 3 | Entanglement recovery via strain-tuning of the QD. (a)** Fine structure splitting (FSS) of the X level for different values of the electric field applied to legs 1 and 4 of the piezoelectric device while varying the value of the electric field applied to legs 2 and 5. The solid lines are given as a guide to the eye. The differently coloured circled points correspond to the curves in the polar plots of panels b and c. **(b)** Polar plot of the distance of the X emission energy from its unperturbed value for two different fields on legs 2 and 5 of the device, while keeping the field value of legs 1 and 4 at 0 kV/cm. The straight arrows highlight the alleged strain direction while the curved arrows highlight the rotation of the polarization angle. **(c)** Same as b but for different values of the field of legs 1 and 4 of the device, while keeping the value of legs 2 and 5 at 6.67 kV/cm. **(d)** Fully entangled fraction, namely the maximum fidelity to a maximally entangled state versus the FSS, for the emitted photon pair of QD2 (green triangles) with higher Purcell factor ($\tau_X = 44(3)\ ps$) and QD4 (pink circles) with lower Purcell factor ($\tau_X = 120(10)\ ps$). The hollow data points correspond to the same measurements corrected for the non-zero value of the $g^{(2)}(0)$. The solid and dashed lines are fits of the data, raw and $g^{(2)}$-corrected respectively, using a simplified model of the FEF, see text. The black star point in the magnified inset is obtained by reducing the laser pulse length to 1.9(3) ps. **(e)** Reconstructed density matrix at the highest fully entangled fraction of QD2 of panel d.

The graph in Fig. 3d shows the FEF of the two-photon state versus the FSS for two QDs (QD2 and QD4) featuring two different values of the lifetime for the X transition, i.e., QD2 with 44(3) ps and QD4 with 120(10) ps, and excited with a 5(1) ps laser pulse length (see section S3 of the Supplementary Information). As expected, the QD with a shorter (longer) lifetime shows a level of entanglement that varies slowly (rapidly) with the FSS[49]. For the two QDs, we fit the experimental FEF with a simplified model[75]:

$$FEF = \frac{1}{4}\left(1 + k + \frac{2k}{\sqrt{1 + \left(\frac{s\tau_X}{\hbar}\right)^2}}\right)$$

where $\tau_X$ is the lifetime of the X state, $s$ is the magnitude of the FSS, and $k$ is a parameter that takes into account decoherence processes and multiphoton emission[49,75]. The value of the X lifetime obtained for the QD2 curve fit is 51(5) ps which is in good agreement with the measured lifetime of 44(3) ps. The fitted value for QD4, i.e., the longer lifetime curve, is 164(9) ps, which is significantly larger than the measured one of 120(10) ps. The reason for this discrepancy is not clear and may suggest other causes for entanglement degradation such as decoherence, spin noise or other entanglement degrees of freedom not taken into account in the model that might impact a longer lifetime QD more, e.g., a relatively large lifetime means that the system has more time to dephase. The values for $k$ are $k = 0.892(6)$ for QD2 and $k = 0.898(7)$ for QD4. The maximum raw value for the FEF is 0.93(1) and is reached for a FSS of 0.2(2) µeV, see inset of Fig. 3d. The corresponding density matrix is shown in Fig. 3e. We can take into account the effect of the non-zero values of the $g^{(2)}(0)$ in the calculation of the FEF values, by removing the counts stemming from multiphoton emission from the coincidences[80]. The corrected value for the FEF at the same FSS value is 0.95(1) which corresponds to a concurrence of 0.90(2). The $g^{(2)}$-corrected data points for the FEF are plotted in the same graph in Fig. 3d as hollow points and fitted with the same model. As expected, the lifetime values stay the same while the values for the fraction $k = 0.920(6)$ is higher and the same for both QDs.

An improved value for the corrected FEF, i.e., 0.96(1), see star point in the inset of Fig. 3d, is measured after reducing the AC-Stark induced shift on the X level from the residual laser pulse during the emission of photons[51,81] by decreasing the laser pulse temporal length to 1.9(3) ps.

## Discussion

To summarize, in this work we demonstrate for the first time a device that combines Purcell-enhanced QDs and a piezoelectric actuator that tunes them for maximizing the degree of polarization entanglement of the emitted photon pairs. More specifically, we show the generation of photon pairs with an entanglement fidelity as high as 0.96(1) and, simultaneously, with an extraction efficiency up to 0.69(4). While these values taken individually do not surpass the best figures of merit that can be found in the literature[27,28,30], our work sets the state of the art for a deterministic source of non-classical light that optimizes both brightness and entanglement, see Table 1. This is highly relevant for real-life application in quantum communication, and in particular for entanglement-based quantum key distribution (E-QKD). The successful implementation of an E-QKD protocol depends on the evaluation of the quantum bit error rate (QBER) and the Bell parameter $S$[82] which both strongly depend on the entanglement fidelity of the photon source used[45]. We envisage that further improvements of the device concept we propose here could be used to exceed[83] the key rates achievable with ideal SPDC sources[45,84] in an E-QKD protocol. A conceptually simple (but technologically challenging) improvement is the integration of a diode-like structure[85,61] onto the micro-machined piezoelectric actuator by changing the geometry of the photonic structure to allow for electrical contacts in the vicinity of the QD via small bridges between the CBR rings[86]. This would allow the application of an electric field across the QDs to counteract possible decoherence mechanisms related to charge noise and thus boost the degree of entanglement to unity values[49]. Moreover, allowing controlled charge tunnelling into/out of the QDs would also enable blinking suppression[87], which would push the flux of entangled photons impinging on the detectors up to a factor of about 3 (we estimate 31 Mcps at the single photon detector for the QDs used here), i.e., values close to the record obtained for single photons[61]. The Purcell enhancement plays a major role in relaxing the demand for low FSS. By looking at Fig 3d, the shorter lifetime QD (green curve) exhibits a corrected fidelity above 0.90 for up to 4 µeV of FSS while at the same time showing an emission energy shift of approximately 40 µeV (0.02 nm).

The reduced tuning range shown in this sample is arguably due to the 50 µm thick GaAs carrier introduced to stabilize the membrane. CBRs structures on a monolithic piezoelectric[65] the achieved tuning range is 1.74 µeV/V, see Table 1, but with a 30 µm thick GaAs carrier. Previous works[36] using a 100 nm thick semiconductor membrane with no GaAs carrier show a tuning range which is more than two orders of magnitude higher (16 µeV/V) than the one achieved in this work against a simulated maximum range of 160 µeV/V[55]. Getting tuning ranges closer to these values could be achieved by thinning down the GaAs carrier to a thickness in the order of hundreds of nm at the cost of introducing the technological challenge of providing mechanically stable membranes capable of enduring all the processing steps involved. We can, however, make more quantitative calculations on the advantage of having a reduced amount of tuning compared to no tuning at all. If we consider the energy distribution of the QDs emission energy in these samples (a normal distribution with 5 meV of standard deviation), the probability of finding a QD which has an emission energy within a distance equal to 1% of its linewidth from a specific line, e.g. the Rb absorption, is around 1 to 100'000. With a 90neV/V tuning range, considering the maximum voltage range sustainable by the piezoelectric substrates (up to ±600 V), the amount of QDs that can be tuned in resonance with a specific line is ~1%, with an improvement of almost 3 orders of magnitude.

Further refinements of the device concept would also enable the use of QDs in other quantum communication applications that need indistinguishable photons, such as entanglement swapping, quantum repeaters, and in general, multi-node quantum networks. These protocols will require levels of indistinguishability beyond the 70% demonstrated in this work. While the use of an electric field will certainly help in stabilizing the charge environment and reduce spectral wandering[88], boosting the indistinguishability to near unity values requires overcoming the time-correlation between the photon pairs generated during the biexciton cascade. A possible solution to alleviate this hurdle would be the use of a cavity that exploits the Purcell effect to engineer the ratio between the exciton and biexciton lifetime[29]. For example, a ratio of 3, a value that can be easily reached in our current device structure, would result in a theoretical photon indistinguishability of up to 0.86. Larger values could be in principle achieved by improving the cavity quality factor while keeping the same broadband extraction efficiency[27]. All these steps will certainly require additional technological advances. However, the efforts are justified as the development of a deterministic source of entangled photons that optimize brightness, degree of entanglement, and indistinguishability would mark the departure from a pioneering phase that is lasting for more than 20 years and would finally open the path towards the exploitation of QDs in real-world applications.

**Table 1 |** Comparison between the figures of merit of this device and state of the art for different device architectures.

| | Extraction efficiency[(1)] | Multiphoton probability | Indistinguishability | Strain tunability | Entanglement fidelity | Tuning of the FSS |
|---|---|---|---|---|---|---|
| This work | 0.69(4) | 0.012(1) | 0.71(1) | 90 neV/V | 0.96(1) | Yes |
| CBRs | 0.85(3)[27] | 0.001(1)[27] | 0.903(3)[27] | - | 0.90(1)[28] | No |
| CBRs on mono. piezo[65] | 0.104 | 0.0015(5) | 0.22(2) | 1.74 µeV/V | - | No |
| Planar cavity[3] | 0.07[41] | 0.008(2)[32] | 0.93(7)[30] | 16 µeV/V[(2)] | 0.98(1)[32] | Yes |
| Broadband antenna[26] | 0.65(4) | 0.002(2) | - | - | 0.90(3) | No |
| Membrane on chip[37] | - | - | - | 120 µeV/V | 0.733(75) | Yes |

[(1)] Single photon
[(2)] Measured on a 100nm thin membrane without DBR planar cavity[36]

## Methods

**Semiconductor QD Sample Structure.** A sacrificial layer of $Al_{0.75}Ga_{0.25}As$ is first grown on a 350 µm thick GaAs (001) commercial substrate in a MBE system. Then, the membrane containing the QDs is grown by first depositing a 4 nm thick layer of GaAs to protect the AlGaAs matrix from oxidation, followed by a first 69 nm thick layer of $Al_{0.33}Ga_{0.67}As$. The QDs are obtained by evaporating Al on the surface to form droplets drilling highly symmetric nanoholes on the surface[54]. The holes are then filled with a 1.5 nm thick layer of GaAs and capped with another 70 nm thick layer of $Al_{0.33}Ga_{0.67}As$ for a total thickness of the membrane of roughly 140 nm. The structure is then protected with another 4 nm thick layer of GaAs bringing the total thickness to 148 nm.

**Cryogenic microscope setup.** To record the positions of QDs for the deterministic fabrication of the CBRs we employ cryogenic imaging using two light sources simultaneously. A blue light emitting diode (LED) (central wavelength of 470 nm) excites QDs above-band gap while an infrared (IR) LED (central wavelength of 810 nm) illuminates the sample located in a liquid-He continuous-flow cryostat, optically accessed using a 0.85 NA glass-corrected objective through a 200 µm thick window. An image of the spatially resolved PL signal and the reference markers is formed on a cost-effective CMOS camera. High-resolution images are acquired with low gain and an exposure time of 1 s and are numerically processed with a Python script that fits the reference markers with straight lines and the QD emission spots with 2D-Gaussian functions. The size of QD spots in the image is close to the diffraction limit. Repeating the detection process of single QDs in 30 different images of the same marker field yields statistical information on the position accuracy, with the most common value for the standard deviation below 15 nm. A more detailed description of the cryogenic imaging setup and the numerical methods is provided in section S1 of the Supplementary Information.

**Photoluminescence Setup.** The processed sample is mounted on a sample holder and the six legs of the micromachined piezoelectric substrate are contacted with Manganin wires. The sample holder is thermally connected to the cold finger of a closed cycle He cryostat which is equipped with electrical feedthroughs for the application of high voltages to the micromachined piezoelectric actuator. A 0.5 NA aspheric lens is used to focus the laser light and collect the photoluminescence (PL) signal. The sample is cooled down to 5 K and is excited with a mode-locked pulsed Ti:Sapphire laser. The 140 fs long laser pulses are narrowed in energy with a 4-f pulse shaper to

5(1) ps temporal width. The pulse shaper also allows for the fine-tuning of the central wavelength of the pulse and changing the laser temporal pulse width down to 1.9(3) ps. The QD is excited with a TPE scheme[15,17] where the laser energy is tuned to half the energy difference of the ground state-XX transition. In this way, the QD is resonantly excited directly to the XX state by absorbing two photons from the laser. A white halogen lamp is focused on the QD to neutralize the charge environment and allow for the TPE[74]. The laser signal reflected from the sample is filtered with a set of three volume Bragg grating filters. The PL signal emitted by the QD is analysed with a 750 mm spectrometer equipped with 300, 1200, and 1800 g/mm gratings and a liquid-nitrogen-cooled CCD camera.

**Fine Structure Splitting Measurement.** The FSS of the X state is measured by placing in the path to the spectrometer a half-wave plate (HWP) and a linear polarizer. The polarization-resolved spectra of the QD emission are collected at each step of the rotation of the HWP using the 1800 g/mm grating. The half-amplitude of a sinusoidal fit of the energy difference between the X and XX line returns the magnitude of the FSS of the X level[89] with sub-µeV accuracy.

**Lifetime Measurements.** The X and XX emission lines of the QD are selected with the 300g/mm grating of the spectrometer and the signal is sent to a low-time jitter (70 ps FWHM) SPAD. The signal of the SPAD is sent to a time correlator with a time jitter of 8 ps. Here, a start-stop histogram is created using the TTL signal from a photodiode inside the laser head as a time reference. The instrument response function is obtained by sending the 5(1) ps long laser pulse along the same path. To extract the values of the lifetimes, we perform a fit[90] by the convolution of the IRF with the exponential decay expected from a simple rate equation model of the radiative cascade. For the XX decay, a single exponential is used for the fit, while for the X decay, the fit is done with an exponential decay preceded by an exponential rise with a lifetime equal to the XX decay time. The error on the lifetime is given by computing the $\chi^2$ surface and taking the confidence interval enclosed in a 5% increase of the $\chi^2$.

**Second-order Correlation Measurements.** The signal coming from either the X or the XX transition is separated from the beam path with a volume Bragg grating mirror and sent to a Hanbury-Brown and Twiss[91] setup. Here, a 50:50 fibre beam splitter sends the incoming photons to two SPADs with a time jitter of about 350 ps. The signal from the SPADs is sent to the time correlator that creates a histogram of the coincidences from the two detectors. The values for the $g^{(2)}(0)$ are calculated by normalizing the counts at the 0-time delay to the counts of the side peaks corresponding to consecutive laser pulses.

**Hong-Ou-Mandel Interference Visibility.** To make photons from two consecutive laser pulses interfere, the pulses of the laser are first doubled with an unbalanced Mach-Zehnder interferometer built with a 1.8 ns time difference between the two arms. The same delay difference is then introduced between the arms of a second Mach-Zehnder interferometer in the path of the PL signal. The histogram of the coincidences is collected from the acquisition events of two SPADs at the exit ports of the last fibre beam splitter. The polarization of the photons impinging on the second beam splitter is selected with a linear polarizer and adjusted with a three-pad fibre polarization manual controller on each input arm of the fibre beam splitter.

**2-photon Density Matrix Reconstruction.** The density matrix is reconstructed by performing polarization-dependent cross-correlation measurements[92–94] between X and XX photons coupled into single-mode fibres. The matrix is reconstructed from a set of 36 measurements associated with different combinations of polarization bases and using a maximum likelihood method. The error bars on each point of the fidelity are obtained with a Monte Carlo simulation consisting of 2000 runs, assuming a Poissonian error on the coincidence counts.

## References

1. Awschalom, D. *et al.* Development of Quantum Interconnects (QuICs) for Next-Generation Information Technologies. *PRX Quantum* **2**, 017002 (2021).

2. Vajner, D. A., Rickert, L., Gao, T., Kaymazlar, K. & Heindel, T. Quantum Communication Using Semiconductor Quantum Dots. *Adv. Quantum Technol.* **5**, 2100116 (2022).

3.	Lu, C.-Y. & Pan, J.-W. Quantum-dot single-photon sources for the quantum internet. *Nat. Nanotechnol.* **16**, 1294–1296 (2021).

4.	Pelucchi, E. *et al.* The potential and global outlook of integrated photonics for quantum technologies. *Nat. Rev. Phys.* **4**, 194–208 (2021).

5.	Uppu, R., Midolo, L., Zhou, X., Carolan, J. & Lodahl, P. Quantum-dot-based deterministic photon–emitter interfaces for scalable photonic quantum technology. *Nat. Nanotechnol.* **16**, 1308–1317 (2021).

6.	Wehner, S., Elkouss, D. & Hanson, R. Quantum internet: A vision for the road ahead. *Science* **362**, eaam9288 (2018).

7.	Pan, J.-W. *et al.* Multiphoton entanglement and interferometry. *Rev. Mod. Phys.* **84**, 777–838 (2012).

8.	Flamini, F., Spagnolo, N. & Sciarrino, F. Photonic quantum information processing: a review. *Rep. Prog. Phys.* **82**, 016001 (2019).

9.	Zhong, H.-S. *et al.* 12-Photon Entanglement and Scalable Scattershot Boson Sampling with Optimal Entangled-Photon Pairs from Parametric Down-Conversion. *Phys. Rev. Lett.* **121**, 250505 (2018).

10.	Meraner, S. *et al.* Approaching the Tsirelson bound with a Sagnac source of polarization-entangled photons. *SciPost Phys.* **10**, 017 (2021).

11.	Takeoka, M., Jin, R.-B. & Sasaki, M. Full analysis of multi-photon pair effects in spontaneous parametric down conversion based photonic quantum information processing. *New J. Phys.* **17**, 043030 (2015).

12.	Jöns, K. D. *et al.* Bright nanoscale source of deterministic entangled photon pairs violating Bell's inequality. *Sci. Rep.* **7**, 1700 (2017).

13.	Schweickert, L. *et al.* On-demand generation of background-free single photons from a solid-state source. *Appl. Phys. Lett.* **112**, 093106 (2018).

14.	Hanschke, L. *et al.* Quantum dot single-photon sources with ultra-low multi-photon probability. *Npj Quantum Inf.* **4**, 43 (2018).

15.	Jayakumar, H. *et al.* Deterministic Photon Pairs and Coherent Optical Control of a Single Quantum Dot. *Phys. Rev. Lett.* **110**, 135505 (2013).

16.	He, Y.-M. *et al.* On-demand semiconductor single-photon source with near-unity indistinguishability. *Nat. Nanotechnol.* **8**, 213–217 (2013).

17.	Müller, M., Bounouar, S., Jöns, K. D., Glässl, M. & Michler, P. On-demand generation of indistinguishable polarization-entangled photon pairs. *Nat. Photonics* **8**, 224–228 (2014).

18.	Aharonovich, I. & Neu, E. Diamond Nanophotonics. *Adv. Opt. Mater.* **2**, 911–928 (2014).

19.	Turunen, M. *et al.* Quantum photonics with layered 2D materials. *Nat. Rev. Phys.* **4**, 219–236 (2022).

20. Drawer, J.-C. *et al.* Ultra-bright single photon source based on an atomically thin material. Preprint at https://doi.org/10.48550/arXiv.2302.06340 (2023).

21. Lodahl, P. Quantum-dot based photonic quantum networks. *Quantum Sci. Technol.* **3**, 013001 (2018).

22. Schimpf, C. *et al.* Quantum dots as potential sources of strongly entangled photons: Perspectives and challenges for applications in quantum networks. *Appl. Phys. Lett.* **118**, 100502 (2021).

23. Akopian, N. *et al.* Entangled Photon Pairs from Semiconductor Quantum Dots. *Phys. Rev. Lett.* **96**, (2006).

24. Dousse, A. *et al.* Ultrabright source of entangled photon pairs. *Nature* **466**, 217–220 (2010).

25. Versteegh, M. A. M. *et al.* Observation of strongly entangled photon pairs from a nanowire quantum dot. *Nat. Commun.* **5**, 5298 (2014).

26. Chen, Y., Zopf, M., Keil, R., Ding, F. & Schmidt, O. G. Highly-efficient extraction of entangled photons from quantum dots using a broadband optical antenna. *Nat. Commun.* **9**, (2018).

27. Liu, J. *et al.* A solid-state source of strongly entangled photon pairs with high brightness and indistinguishability. *Nat. Nanotechnol.* **14**, 586–593 (2019).

28. Wang, H. *et al.* On-Demand Semiconductor Source of Entangled Photons Which Simultaneously Has High Fidelity, Efficiency, and Indistinguishability. *Phys. Rev. Lett.* **122**, 113602 (2019).

29. Schöll, E. *et al.* Crux of Using the Cascaded Emission of a Three-Level Quantum Ladder System to Generate Indistinguishable Photons. *Phys. Rev. Lett.* **125**, 233605 (2020).

30. Huber, D. *et al.* Highly indistinguishable and strongly entangled photons from symmetric GaAs quantum dots. *Nat. Commun.* **8**, 15506 (2017).

31. Huwer, J. *et al.* Quantum-Dot-Based Telecommunication-Wavelength Quantum Relay. *Phys. Rev. Appl.* **8**, 024007 (2017).

32. Huber, D. *et al.* Strain-Tunable GaAs Quantum Dot: A Nearly Dephasing-Free Source of Entangled Photon Pairs on Demand. *Phys. Rev. Lett.* **121**, 033902 (2018).

33. Hopfmann, C. *et al.* Maximally entangled and gigahertz-clocked on-demand photon pair source. *Phys. Rev. B* **103**, 075413 (2021).

34. Gurioli, M., Wang, Z., Rastelli, A., Kuroda, T. & Sanguinetti, S. Droplet epitaxy of semiconductor nanostructures for quantum photonic devices. *Nat. Mater.* **18**, 799–810 (2019).

35. Trotta, R. *et al.* Nanomembrane Quantum-Light-Emitting Diodes Integrated onto Piezoelectric Actuators. *Adv. Mater.* **24**, 2668–2672 (2012).

36. Trotta, R. *et al.* Wavelength-tunable sources of entangled photons interfaced with atomic vapours. *Nat. Commun.* **7**, 10375 (2016).

37. Chen, Y. *et al.* Wavelength-tunable entangled photons from silicon-integrated III–V quantum dots. *Nat. Commun.* **7**, 10387 (2016).

38. Ou, W. *et al.* Strain Tuning Self-Assembled Quantum Dots for Energy-Tunable Entangled-Photon Sources Using a Photolithographically Fabricated Microelectromechanical System. *ACS Photonics* **9**, 3421–3428 (2022).

39. Reindl, M. *et al.* All-photonic quantum teleportation using on-demand solid-state quantum emitters. *Sci. Adv.* **4**, eaau1255 (2018).

40. Nilsson, J. *et al.* Quantum teleportation using a light-emitting diode. *Nat. Photonics* **7**, 311–315 (2013).

41. Basso Basset, F. *et al.* Entanglement Swapping with Photons Generated on Demand by a Quantum Dot. *Phys. Rev. Lett.* **123**, 160501 (2019).

42. Zopf, M. *et al.* Entanglement Swapping with Semiconductor-Generated Photons Violates Bell's Inequality. *Phys. Rev. Lett.* **123**, 160502 (2019).

43. Schwartz, I. *et al.* Deterministic generation of a cluster state of entangled photons. *Science* **354**, 434–437 (2016).

44. Cogan, D., Su, Z.-E., Kenneth, O. & Gershoni, D. Deterministic generation of indistinguishable photons in a cluster state. *Nat. Photonics* **17**, 324–329 (2023).

45. Basso Basset, F. *et al.* Quantum key distribution with entangled photons generated on demand by a quantum dot. *Sci. Adv.* **7**, (2021).

46. Schimpf, C. *et al.* Quantum cryptography with highly entangled photons from semiconductor quantum dots. *Sci. Adv.* **7**, eabe8905 (2021).

47. Ursin, R. *et al.* Entanglement-based quantum communication over 144 km. *Nat. Phys.* **3**, 481–486 (2007).

48. Acín, A. *et al.* Device-Independent Security of Quantum Cryptography against Collective Attacks. *Phys. Rev. Lett.* **98**, 230501 (2007).

49. Hudson, A. J. *et al.* Coherence of an Entangled Exciton-Photon State. *Phys. Rev. Lett.* **99**, 266802 (2007).

50. Schimpf, C. *et al.* Hyperfine interaction limits polarization entanglement of photons from semiconductor quantum dots. *Phys. Rev. B* **108**, L081405 (2023).

51. Basso Basset, F. *et al.* Signatures of the Optical Stark Effect on Entangled Photon Pairs from Resonantly-Pumped Quantum Dots. Preprint at https://doi.org/10.48550/arXiv.2212.07087 (2023).

52. Fognini, A. *et al.* Dephasing Free Photon Entanglement with a Quantum Dot. *ACS Photonics* **6**, 1656–1663 (2019).

53. Bayer, M. *et al.* Electron and Hole g Factors and Exchange Interaction from Studies of the Exciton Fine Structure in In 0.60 Ga 0.40 As Quantum Dots. *Phys. Rev. Lett.* **82**, 1748–1751 (1999).

54. Huo, Y. H., Rastelli, A. & Schmidt, O. G. Ultra-small excitonic fine structure splitting in highly symmetric quantum dots on GaAs (001) substrate. *Appl. Phys. Lett.* **102**, 152105 (2013).

55. Trotta, R., Martín-Sánchez, J., Daruka, I., Ortix, C. & Rastelli, A. Energy-Tunable Sources of Entangled Photons: A Viable Concept for Solid-State-Based Quantum Relays. *Phys. Rev. Lett.* **114**, 150502 (2015).

56. Wang, J., Gong, M., Guo, G.-C. & He, L. Towards Scalable Entangled Photon Sources with Self-Assembled InAs/GaAs Quantum Dots. *Phys. Rev. Lett.* **115**, 067401 (2015).

57. Trotta, R. *et al.* Universal Recovery of the Energy-Level Degeneracy of Bright Excitons in InGaAs Quantum Dots without a Structure Symmetry. *Phys. Rev. Lett.* **109**, 147401 (2012).

58. Senellart, P., Solomon, G. & White, A. High-performance semiconductor quantum-dot single-photon sources. *Nat. Nanotechnol.* **12**, 1026–1039 (2017).

59. Ma, Y., Kremer, P. E. & Gerardot, B. D. Efficient photon extraction from a quantum dot in a broad-band planar cavity antenna. *J. Appl. Phys.* **115**, 023106 (2014).

60. Purcell, E. M., Torrey, H. C. & Pound, R. V. Resonance Absorption by Nuclear Magnetic Moments in a Solid. *Phys. Rev.* **69**, 37–38 (1946).

61. Tomm, N. *et al.* A bright and fast source of coherent single photons. *Nat. Nanotechnol.* **16**, 399–403 (2021).

62. Schliwa, A., Winkelnkemper, M. & Bimberg, D. Few-particle energies versus geometry and composition of In x Ga 1 − x As / GaAs self-organized quantum dots. *Phys. Rev. B* **79**, 075443 (2009).

63. Ginés, L. *et al.* High Extraction Efficiency Source of Photon Pairs Based on a Quantum Dot Embedded in a Broadband Micropillar Cavity. *Phys. Rev. Lett.* **129**, 033601 (2022).

64. Su, M. Y. & Mirin, R. P. Enhanced light extraction from circular Bragg grating coupled microcavities. *Appl. Phys. Lett.* **89**, 033105 (2006).

65. Moczała-Dusanowska, M. *et al.* Strain-Tunable Single-Photon Source Based on a Circular Bragg Grating Cavity with Embedded Quantum Dots. *ACS Photonics* **7**, 3474–3480 (2020).

66. Hardy, A., Welch, D. F. & Streifer, W. Analysis of second-order gratings. *IEEE J. Quantum Electron.* **25**, 2096–2105 (1989).

67. Davanço, M., Rakher, M. T., Schuh, D., Badolato, A. & Srinivasan, K. A circular dielectric grating for vertical extraction of single quantum dot emission. *Appl. Phys. Lett.* **99**, 041102 (2011).

68. Rickert, L., Kupko, T., Rodt, S., Reitzenstein, S. & Heindel, T. Optimized designs for telecom-wavelength quantum light sources based on hybrid circular Bragg gratings. *Opt. Express* **27**, 36824 (2019).

69. Kolatschek, S. *et al.* Bright Purcell Enhanced Single-Photon Source in the Telecom O-Band Based on a Quantum Dot in a Circular Bragg Grating. *Nano Lett.* **21**, 7740–7745 (2021).

70. Galal, H. & Agio, M. Highly efficient light extraction and directional emission from large refractive-index materials with a planar Yagi-Uda antenna. *Opt. Mater. Express* **7**, 1634 (2017).

71. Liu, J. *et al.* Cryogenic photoluminescence imaging system for nanoscale positioning of single quantum emitters. *Rev. Sci. Instrum.* **88**, 023116 (2017).

72. Huang, H. *et al.* Electrically-Pumped Wavelength-Tunable GaAs Quantum Dots Interfaced with Rubidium Atoms. *ACS Photonics* **4**, 868–872 (2017).

73. Huber, D. *et al.* Single-particle-picture breakdown in laterally weakly confining GaAs quantum dots. *Phys. Rev. B* **100**, 235425 (2019).

74. Huber, T., Predojević, A., Solomon, G. S. & Weihs, G. Effects of photo-neutralization on the emission properties of quantum dots. *Opt. Express* **24**, 21794 (2016).

75. Rota, M. B., Basset, F. B., Tedeschi, D. & Trotta, R. Entanglement Teleportation With Photons From Quantum Dots: Toward a Solid-State Based Quantum Network. *IEEE J. Sel. Top. Quantum Electron.* **26**, 1–16 (2020).

76. Hong, C. K., Ou, Z. Y. & Mandel, L. Measurement of subpicosecond time intervals between two photons by interference. *Phys. Rev. Lett.* **59**, 2044–2046 (1987).

77. Somaschi, N. *et al.* Near-optimal single-photon sources in the solid state. *Nat. Photonics* **10**, 340–345 (2016).

78. Trotta, R. & Rastelli, A. Engineering of Quantum Dot Photon Sources via Electro-elastic Fields. in *Engineering the Atom-Photon Interaction* (eds. Predojević, A. & Mitchell, M. W.) 277–302 (Springer International Publishing, Cham, 2015).

79. Grondalski, J., Etlinger, D. M. & James, D. F. V. The fully entangled fraction as an inclusive measure of entanglement applications. *Phys. Lett. A* **300**, 573–580 (2002).

80. Neuwirth, J. *et al.* Multipair-free source of entangled photons in the solid state. *Phys. Rev. B* **106**, L241402 (2022).

81. Seidelmann, T. *et al.* Two-Photon Excitation Sets Limit to Entangled Photon Pair Generation from Quantum Emitters. *Phys. Rev. Lett.* **129**, 193604 (2022).

82. Clauser, J. F., Horne, M. A., Shimony, A. & Holt, R. A. Proposed Experiment to Test Local Hidden-Variable Theories. *Phys. Rev. Lett.* **23**, 880–884 (1969).

83. Acín, A., Massar, S. & Pironio, S. Efficient quantum key distribution secure against no-signalling eavesdroppers. *New J. Phys.* **8**, 126–126 (2006).

84. Hošák, R., Straka, I., Predojević, A., Filip, R. & Ježek, M. Effect of source statistics on utilizing photon entanglement in quantum key distribution. *Phys. Rev. A* **103**, 042411 (2021).

85. Salter, C. L. *et al.* An entangled-light-emitting diode. *Nature* **465**, 594–597 (2010).

86. Buchinger, Q., Betzold, S., Höfling, S. & Huber-Loyola, T. Optical properties of circular Bragg gratings with labyrinth geometry to enable electrical contacts. *Appl. Phys. Lett.* **122**, 111110 (2023).

87. Schimpf, C., Manna, S., Covre da Silva, S. F., Aigner, M. & Rastelli, A. Entanglement-based quantum key distribution with a blinking-free quantum dot operated at a temperature up to 20 K. *Adv. Photonics* **3**, (2021).

88. Troiani, F. Entanglement swapping with energy-polarization-entangled photons from quantum dot cascade decay. *Phys. Rev. B* **90**, 245419 (2014).

89. Young, R. J. *et al.* Bell-Inequality Violation with a Triggered Photon-Pair Source. *Phys. Rev. Lett.* **102**, 030406 (2009).

90. Preus, S. DecayFit - Fluorescence Decay Analysis Software 1.4, FluorTools, www.fluortools.com. FluorTools (2023).

91. Hanbury Brown, R. & Twiss, R. Q. A Test of a New Type of Stellar Interferometer on Sirius. *Nature* **178**, 1046–1048 (1956).

92. Ježek, M., Fiurášek, J. & Hradil, Z. Quantum inference of states and processes. *Phys. Rev. A* **68**, 012305 (2003).

93. Hradil, Z., Řeháček, J., Fiurášek, J. & Ježek, M. 3 Maximum-Likelihood Methods in Quantum Mechanics. in *Quantum State Estimation* (eds. Paris, M. & Řeháček, J.) 59–112 (Springer, Berlin, Heidelberg, 2004).

94. Altepeter, J. B. *et al.* Experimental Methods for Detecting Entanglement. *Phys. Rev. Lett.* **95**, 033601 (2005).

# Acknowledgements

RT acknowledges the European Research Council (ERC) under the European Union's Horizon 2020 Research and Innovation Program under Grant Agreement No. 679183 (SPQRel), the European Union's Horizon 2020 (2014–2020) under Grant Agreement No. 731473 (QuantERA project HYPER-U-P-S No. 8C18002) and the MUR (Ministero dell'Università e Ricerca) through the PNRR MUR Project PE0000023-NQSTI. RT and AR acknowledge the European Union's Horizon 2020 Research and Innovation Program under Grant Agreement No. 899814 (Qurope), the QuantERA II Programme that has received funding from the European Union's Horizon 2020 Research and Innovation Programme under Grant Agreement No. 101017733 via the project QD-E-QKD and the FFG (grant No. 891366). AR also acknowledges the European Union's Horizon 2020 Research and Innovation Program under Grant Agreements No. 871130 (Ascent+), the Austrian Science Fund (FWF) via the Research Group FG5, P 29603, I 4380, I 3762, the Linz Institute of Technology (LIT), and the LIT Secure and Correct Systems Lab, supported by the State of Upper Austria. JN acknowledges MUR via Project PRIN 2017 Taming complexity via Quantum Strategies a Hybrid Integrated Photonic approach (QUSHIP) Id. 2017SRNBRK. THL acknowledges funding from the Federal Ministry of Education and Research (BMBF) through the Quantum Futur (FKZ: 13N16272) initiative. MJ acknowledges the Czech Science Foundation (grant no. 21-18545S). NH acknowledges the European Union's 2020 research and innovation program (CSA-Coordination and support action, H2020-WIDESPREAD-2020-5) under grant agreement No. 951737 (NONGAUSS) and under Grant Agreement No. 101017733 via the project QD-E-QKD, and Palacky University (grants no. IGA-PrF-2022-005, IGA-PrF-2023-006). XY acknowledges support from the National Natural Science Foundation of China (NSFC 12104090),

"the Fundamental Research Funds for the Central Universities" and "Zhishan" Scholars Programs of Southeast University. The authors thank B. U. Lehner, T. Oberleitner, and C. Schimpf for fruitful discussions, as well as U. Kainz, A. Halilovic and A. Schwarz from JKU, A. Miriametro from Sapienza University of Rome, and S. Kuhn from University of Würzburg for technical assistance.